# AI-Enhanced Intelligent NIDS Framework: Leveraging Metaheuristic Optimization for Robust Attack Detection and Prevention

Maryam Mahdi Alhusseini [a*], (Member, IEEE), Mohammad-Reza Feizi-Derakhshi [b]

[a] Information and Communication Technology Department, Middle Technical University, Baghdad, Iraq,
[b] Computerized Intelligence Systems Laboratory, Department of Computer Engineering, University of Tabriz, Tabriz, Iran

Corresponding Author: Maryam Mahdi Alhusseini, Email [*]: mariammahdi@mtu.edu.iq

**Abstract**
In today's rapidly evolving digital landscape, safeguarding network infrastructures against cyberattacks has become a critical priority. This research presents an innovative AI-driven real-time intrusion detection framework designed to enhance network security, particularly in Wireless Sensor Networks (WSNs), Cloud Computing (CC), and Internet of Things (IoT). environments. The system employs classical machine learning models, Logistic Regression, Decision Tree, and K-Nearest Neighbors, optimized through the novel Energy Valley Optimization (EVO) method using the NSL-KDD dataset. Feature selection significantly reduced the number of input features from 42 to 18, while maintaining strong detection capabilities. The proposed system achieved 98.95% accuracy with Decision Tree, 98.47% with K-Nearest Neighbors, and 88.84% with Logistic Regression. Moreover, high precision, recall, and F1-scores were attained across all classifiers while substantially reducing training and testing times, making the framework highly suitable for real-time applications. To ensure fair detection across diverse attack types, dataset balancing via downsampling was applied to address class imbalance challenges. This investigation focuses on the significance of advancing IDSs in cloud computing and WSNs. Overall, this work advances secure communications by delivering a scalable, low-latency, and high-accuracy intrusion detection solution aligned with the latest trends in artificial intelligence, cybersecurity, and real-time digital networks.

## 1   INTRODUCTION

This article addresses the pressing need for enhancing Network Intrusion Detection Systems (NIDS) within the domains of Wireless Sensor Networks (WSNs), Cloud Computing, and the Internet of Things (IoT). Integrating cloud technology with wired and wireless sensor networks is crucial in various applications such as transportation, warfare, education, healthcare, and agriculture (Figure 1). The primary problem at hand is the detection of intrusions with high accuracy while minimizing false positives. Conventional IDSs often suffer from high false positive rates, leading to inefficient resource utilization and potentially overlooking genuine threats. Thus, the central aim of this paper is to propose a novel hybrid IDS model that leverages cutting-edge methodologies to achieve superior detection accuracy and reduce false positives (FP).

The domain of cybersecurity is a major problem area for security staff, given the increasing volume and variety, and frequency of cyber-attacks and evil activities. Learning from high-dimensional data with a large number of features is a difficult challenge in machine learning, as high-dimensional data can challenge model efficacy, increase computational overhead, and complicate feature choice. It makes the machine learning approaches struggle when dealing with numerous types of input features, which is a major obstacle to researchers. Preparation of the data is critical for the machine learning algorithms. Feature selection is a common methodology in data preprocessing required in machine learning [1].

Feature selection (FS) Strategy is the process of identifying relevant attributes or a potential subset of attributes. Evaluation criteria are used to obtain an optimal subset of attributes. In high-dimensional data (where the number of samples is much smaller than the number of attributes), identifying the appropriate subset of attributes is a challenging endeavor [1]. The evolutionary algorithms enhance problem-solving across various domains. The Forest Optimization Algorithm (FOA) was initially developed for continuous search but has been adapted for discrete search in feature selection [2]. The imperative for proficient management of high-dimensional data has resulted in the creation of feature selection techniques to remove irrelevant and redundant attributes, hence enhancing generalization and reducing noise.

Metaheuristic algorithms are crucial for feature selection in high-dimensional datasets as they can navigate intricate solution spaces and circumvent local optima. Energy valley Optimizer [3], Particle swarm optimization (PSO), genetic algorithms (GA), and ant colony optimization (ACO) are popular feature selection techniques due to their efficiency and versatility. Hacking attacks are among the most common types of cyberattacks [4]. Meta-optimization algorithms are new types of optimizers that stand out and claim that they can do simulated annealing and ant colony optimization, as well as the bee algorithms. These optimizers have been used in many works focused on the improvement of classification systems through feature selection [5]. Cybersecurity challenges are a fundamental requirement of modern computer systems, having recently acquired considerable importance due to the substantial increase in both the frequency and intensity of malicious attacks [6]. Intrusion Detection Systems (IDSs) are essential for alleviating these risks [7]. Many protection measures, IDS, are needed to detect attackers and harmful programs in the cloud computing environment, which relies heavily on the Internet [8]. Machine learning techniques in IDS (with hyperparameter tuning) assisted by the Energy valley optimizer (EVO) can increase the precision and accuracy of anomaly detection in the system [9]. Leveraging advanced optimization techniques, such as the Energy Valley Optimizer (EVO) for feature selection in an intrusion detection system (IDS), improves detection accuracy and ensures smooth network operations, just as congestion control algorithms improve TCP performance. An intrusion detection system (IDS) is a software or hardware solution that monitors internal and external networks or system administrators. The ability of an IDS to detect both known and new attacks makes it an excellent choice for protecting cloud computing [10]. Anomaly detection accuracy is demonstrated by advanced monitoring methods, such as optical fiber sensors (OFS), which are characterized by their sensitivity and accuracy. Intrusion detection systems (IDS) that use machine learning (ML) and the Energy Valley Optimizer (EVO) enhance cybersecurity through effective monitoring and feature selection [3].

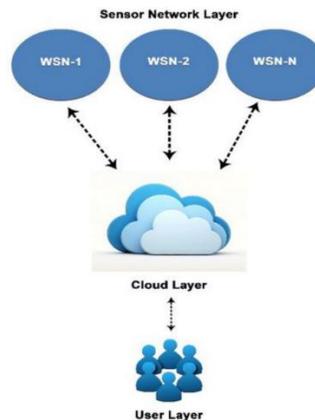

**Figure 1.** Integration of WSNs to the cloud [11]

To address these challenges, the article proposes employing a unique feature selection method, specifically the Energy Valley feature selection algorithm [3], to analyze substantial volumes of potential intrusion data. To assess our novel approach, our proposed approach was evaluated on the Intrusion Detection System (IDS) platform, a sophisticated system employing machine learning techniques, and applied to actual large-scale intrusion data NSL_KDD. Our paper proposes an optimized approach for detecting malicious packets by integrating metaheuristic algorithms into an Intrusion Detection System. The proposed algorithm aims to improve accuracy and precision while reducing space and time complexity by integrating metaheuristic algorithms with existing machine learning classifier techniques. The experimental results demonstrate that this hybrid approach outperforms existing classifiers, making it a promising solution for IDS optimization.

In the preliminary stage of our methodology, the novel Energy Valley Optimizer algorithm (EVO) [3] was employed to generate an optimal feature set using feature selection (FS). This feature selection method significantly enhances the system's capacity to accurately represent data and improves the precision of intrusion detection.

During the second phase, three unique machine learning models, Logistic Regression (LogReg), Decision Tree (D_Tree), and K-Nearest Neighbors (KNN), are employed to attain comprehensive and efficient intrusion detection utilizing data enhanced by our proposed methodology.

The IDS is evaluated by rigorous tests on real-world NSL_KDD datasets across various scenarios, with a total of 6 experiments executed. The system's performance is evaluated based on detection accuracy, Precision, Sensitivity, False Positive Rates (FPR), and computational efficiency.

This article pioneers the use of the Energy Valley Optimizer (EVO) for feature selection within the area of intrusion detection, achieving a significant feature reduction in the number of features, improved detection performance, and real-time operational requirements relevant to networked and embedded systems.

This paper's major contributions are as follows:
- ✓ Introduced the first approach of the Energy Valley Optimizer (EVO) for efficient feature selection, reducing the NSL-KDD dataset's features from 42 to 18 while improving detection performance.
- ✓ Optimized machine learning models (logistic regression, decision tree, KNN) for real-time intrusion detection in networked environments.
- ✓ Achieved superior accuracy, precision, recall, and F1-score compared to existing methods.
- ✓ Demonstrated real-time feasibility with low training and testing times suitable for embedded and communication systems
- ✓ Balanced the dataset using downsampling techniques, ensuring fair and reliable detection across both majority and minority attack classes.

The paper is structured as follows: Section 2 discusses related literature, Section 3 presents the Energy Valley Optimizer for feature selection and IDS, and Section 4 delineates the conclusion of our study. Section 5 addressed future recommendations. This section offers an analysis of the performance and efficacy of IDS in identifying and alleviating cybersecurity threats

## 2 NOVELTY

This investigation pioneers the use of the Energy Valley Optimizer (EVO) for Feature Selection Strategy in intrusion detection, achieving significant feature reduction, improved detection performance, and real-time operational efficiency, specifically tailored for networked wired and wireless

## 3 RELATED WORK

Wireless Sensor networks and Cloud computing (CC) are vulnerable to various security threats and attacks. To ensure their security, an effective Intrusion Detection System (IDS) must be in place to detect these attacks [13]. Traditional IDSs are less effective as these malicious attacks become more intelligent, frequent, and complex, especially in large detection datasets. Intrusion detection systems (IDSs) are critical in safeguarding CC and WSNs [14].

This section analyzes studies on contemporary optimization methods for extensive, high-dimensional datasets utilizing metaheuristic algorithms for feature selection. Furthermore, research has concentrated on machine learning models and optimization techniques in intrusion detection systems (IDS) to improve cybersecurity. Metaheuristics, highlighted by Mirjalili et al. [15], offer effective optimization methods for real-world problems, avoiding local optima and thoroughly exploring complex search spaces. Feature selection (FS) is essential for enhancing efficiency while minimizing memory utilization and computational expenses. Due to the inclusion of unnecessary and redundant features in the data, which complicate the problem and use machine memory, and hinder learning through simulation, feature selection (FS) is advised. Feature selection (FS) is a crucial data mining technique that enhances model learning by removing duplicate and unnecessary characteristics, hence facilitating numerous applications across diverse fields. Sangaiah et al. [16] proposed Hybrid Ant-Bee Colony Optimization (HABCO) to optimize feature selection problems and turn them into an optimization problem. R. Ghanbarzadeh et al. [17] presented an innovative method for network intrusion detection (NIDS) utilizing the Horse Herd Optimization Algorithm (HOA) and Quantum-Inspired Optimization (QIO). H. Nkiama [18] emphasized the significance of feature selection in intrusion detection systems (IDS) to enhance accuracy and performance. The research introduces a recursive feature elimination method using a decision tree classifier to detect and remove irrelevant components. Implementing this methodology on the NSL-KDD dataset yields substantial enhancements in accuracy. Scholars have utilized natural language processing methods to examine large volumes of social data from Twitter in order to gather valuable insights that support intrusion detection and the analysis of cybersecurity threats in intrusion detection systems (IDS) [19]. This research enhances Q-learning by integrating a fuzzy approach to accelerate and refine resource allocation in Intrusion Detection Systems (IDS), which consequently improves compliance with Service Level Agreements (SLA) and Quality of Service (QoS) [20]. Intrusion Detection Systems (IDSs) or Network Intrusion Detection Systems (NIDS) are software utilized to analyze network traffic and detect suspicious activities originating from both internal users and external entities. This paper introduces an innovative "nesting circles" visualization technique aimed at streamlining and enriching the examination of IDS alerts, thereby enhancing the identification of concealed attacks [21]. The NSL-KDD dataset serves as a standard for evaluating intrusion detection systems. These findings underscore the significance of feature selection in the development of successful Intrusion Detection Systems (IDS).

## 3.1 FEATURE SELECTION (FS) STRATEGY

In their work, Eesa et al. [22] introduced a novel feature selection approach for intrusion detection systems (IDS) utilizing the Cuttlefish Optimization Algorithm (CFA), which is inspired by the behavior of squids. Zhao et al. [23] presented a hybrid intrusion detection method utilizing CFS-DE. They employed CFS-DE to identify the optimal feature set for dimensionality reduction. M. R. Feizi Derakhshi et al. [2] demonstrated that evolutionary algorithms optimize problems in several fields. Forest Optimization Algorithm (FOA) was created for continuous search but has been extended for discrete search feature selection. The hybridization of heuristic approaches seeks to integrate the advantages of individual heuristics to enhance outcomes for the optimization issue [24, 25]. P. K. Keserwani et al. [26] introduce a cloud network-specific anomaly-based intrusion detection system. The proposed method employs deep learning (DL) for classification and utilizes hybrid metaheuristics for feature selection. The Grey Wolf Optimization (GWO) and Crow Search Algorithm are utilized for feature selection. This hybrid approach extracts essential cloud network data. Kabir et al. [27] this research present a novel hybrid ant colony optimization (ACO) approach for feature selection (FS), termed ACOFS, which incorporates a neural network. An essential element of this approach is the identification of a smaller subset of significant features. Recent literature trends show an increasing adoption of feature selection techniques, as summarized in Table 1.

**Table 1**. Summary of modern studies in the domain of intrusion detection utilizing *feature selection methodologies*

| Ref. | year | Proposed Approach | Feature Selection Methods | Type of Datasets | Challenges |
|---|---|---|---|---|---|
| [28] | 2017 | NB and NB tree | RF, C4.5 and NB-Tree | KDD99, UNSW-NB15 | Anomaly |
| [29] | 2019 | SVM | IGWEKA'sfilter | KDD99, UNSW-NB15 | Anomaly |
| [30] | 2019 | Soft-maxclassifier | DSAE | NSL_KDD | Anomaly |
| [31] | 2023 | CNN | LSTM | KDD99, NSL-KDD, UNSW-NB15 | Anomaly |

## 4 ENERGY VALLEY OPTIMIZER (EVO)

The Energy Valley Optimizer (EVO), introduced by Mahdi Azizi et al. in January 2023, is an innovative metaheuristic algorithm derived from sophisticated physics concepts [12, 3]. The objective is to enhance the alignment between conceptual and mathematical models, focusing on particle stability about neutron and proton ratios [3]. Advanced monitoring methods such as Optical Fiber Sensors (OFS), recognized for their exceptional sensitivity and accuracy, underscore the significance of precision in anomaly detection. Likewise, Intrusion Detection Systems (IDS) employing Machine Learning (ML) and the Energy Valley Optimizer (EVO) leverage efficient monitoring and feature selection to improve cybersecurity [3]. EVO incorporates a dynamic configuration for exploration and exploitation, employing three novel position vectors throughout its search process [3]. The intricacy, along with its mathematical underpinnings, establishes EVO as a potential instrument for optimization tasks, especially in high-dimensional datasets [3].

### 4.1 MATHEMATICAL FORMULATION OF EVO

This algorithm simulates the process by which unstable particles attempt to reach a stable state by adjusting their neutron-to-proton ratio (N/Z). In nature, unstable particles emit energy and decay into more stable forms, often through alpha, beta, or gamma decay, depending on their levelof instability. EVO models this behavior, where solution candidates (particles) in a search space strive to move towardan optimal solution (stability) by adjusting their parameters [3]. The fundamental concept of EVO is that each particle inside the search space signifies a potential solution, with the algorithm's objective being to identify the most stable (optimal) solution through the iterative evolution of the particles. Each particle modifies its stability level through many phases, including enrichment and decay processes, akin to natural physical occurrences, as shown in Figure 2. The subsequent equations dictate the location updates of particles in the EVO algorithm:

$$X_i^{NEW1} = X_i (X_{BS} (X_t^{i'}))$$

This equation represents the alpha decay process, where a new position vector for the *i-th* particle is generated based on the best particle's stability level $X_{BS}$ and the current position of the particle $X_i$. Here, $x'_i$ is a randomly generated value that controls the degree of adjustment. The goal of this step is to bring the particle closer to the best solution by making small adjustments.

$$X_i^{NEW2} = X_i (X_{NG} (X_t^{i'}))$$

This equation represents the gamma decay process, where the particle's position is updated based on a neighboring particle $X_{NG}$, which has also undergone a decay process. The aim is to encourage collaboration between particles by allowing them to learn from the positions of their neighbors, facilitating convergence toward the optimal solution.

$$X_i^{NEW1} = X_i + (\tau_1 * X_{BS} - \tau_2 * X_{CP})/SL_i$$

This equation models the beta decay process, where a new position is generated based on both the best particle $X_{BS}$ and the center of particles $X_{CP}$. Where $\tau_1$ and $\tau_2$ are weight factors that govern the impact of the best particle and the center of particles, respectively. The tag stability level $SL_i$ of the particle evaluates to the fact that a more unstable particle adjusts with greater scale (larger deviation), as in practice, particles always tend to stabilize.

$$X_i^{NEW2} = X_i + (\tau_3 * X_{BS} - \tau_4 * X_{NG})$$

This equation describes yet another type of beta decay, where a particle's position is aligned towards the best particle and the neighboring particles $X_{NG}$, where $\tau_3$ and $\tau_4$ are weights factor that controls the contribution of the best particle ($x_{best}$) and neighboring particles to spur information sharing between particles to encourage exploration in the algorithm.

Energy Valley Optimizer (EVO) then uses equations like these to represent how the particles search for equilibrium by moving around in the search space. Mechanism of Interaction for each particle with the best particle, neighboring particles, in addition to the center of the swarm, providing both local search and global experience. This study can be extended in the future, as our next step is to use the EVO algorithm with the SNAKE metaheuristic for optimizing the number of neurons and learning rate in deep learning models with even better performance concerning network intrusion detection.

## 4.2 COST FUNCTION IN EVO

The objective function (cost function) utilized in EVO is contingent upon the optimization problem. In the context of Intrusion Detection Systems (IDS), the cost function is generally developed based on classification accuracy, F1-score, or a weighted amalgamation of many performance measures.

*Cost function = $w_1$. (1-Accuracy) + $w_2$. False Positive rate + $w_3$. False Negative Rate*

Where:
W1, W2, and W3 are weighting factors that prioritize different aspects of classification performance.

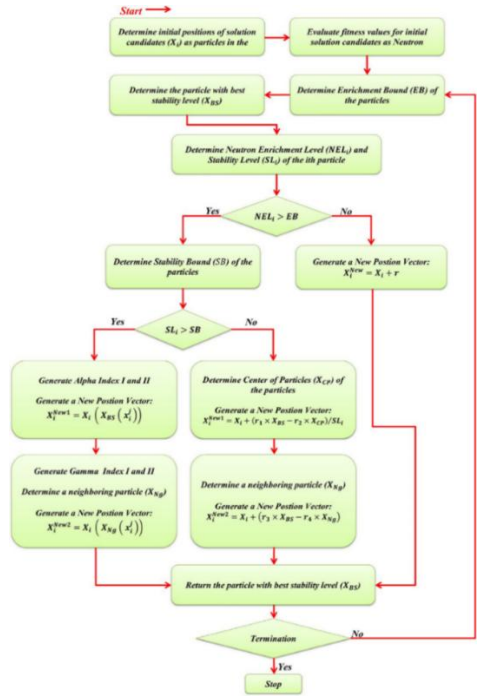

**Figure 2.** Flowchart of the EVO [3]

## 4.3 DATA PROCESSING FOR INTRUSION DETECTION
### 4.3.1 Recent Studies of Intrusion Datasets

Most of the datasets used for intrusion detection system evaluation contain a large amount of feature information. Irrelevant and redundant features may lead to degradation of the performance of the intrusion detection system. feature selection is a common method of selecting important features.

Zhao et al. [23] In high-dimensional datasets, efficient feature selection is essential for pinpointing the most valuable properties and minimizing computing burden. Meta-heuristic optimization techniques have demonstrated potential in overcoming these restrictions by offering a versatile, global search capacity adept at managing diverse feature selection issues. Long et al. [32]. Employed a random attention-based data fusion strategy to eliminate redundant characteristics, followed by the application of a semi-supervised ladder network model for intrusion detection. Mojtahedi et al [33] combined the Whale Optimization Algorithm (WOA) and Genetic Algorithm (GA) for feature selection, and classification by KNN, which got a better result than other previous methods. Li et al. [34] An enhanced krill swarm (KH) optimization algorithm was proposed for feature selection on the NSL-KDD and CICIDS-2017 datasets, retaining an average of 7 and 10.2 features, respectively. This approach effectively eliminates redundant features while ensuring high detection accuracy. To accurately identify cyber threats, Intrusion Detection Systems (IDS) models must analyze large amounts of dynamic data, making feature selection crucial.

The research shows that intrusion detection data is large, complex, and imbalanced between attack and normal data. Many intrusion detection methods require feature selection and data balancing to reduce model training time and improve classification accuracy. This paper's feature selection and data balance strategy ensures detection accuracy. Table 2 Intrusion Detection System literature review overview.

**Table 2.** Summary of Recent Studies on the NSL_KDD Detection Dataset

| Ref. | Datasets | Type of Attacks | Years |
|---|---|---|---|
| [35] | NSL_KDD | Normal, DoS, R2L, U2R, Probing | 2020 |
| [36] | NSL_KDD | Attack, Normal | 2021 |
| [37] | NSL_KDD | Probe, U2R, R2L, DoS | 2022 |
| [38] | NSL_KDD | NORMAL, Probe, U2R, R2L, DoS | 2022 |
| [39] | NSL_KDD | Normal, DoS, R2L, U2R, Probing | 2023 |

### 4.3.2 Proposed Detection Real Datasets NSL_KDD

This dataset has been proposed as a potential solution to address certain limitations present in the KDD'99 dataset. The NSL-KDD dataset improves the KDD Cup 1999 intrusion detection system evaluation dataset by eliminating duplicate instances in both the training set and test set [40]. It is the Network Security Laboratory- Knowledge Discovery in Databases. NSL-KDD is a data set suggested to solve some of the inherent problems of the KDD'99 data set, which are mentioned in. The provided data can be utilized to assist researchers in conducting comparisons between different intrusion detection methods, IDS [40]. One of the most important deficiencies in the KDD data set is the huge number of redundant records, which causes the learning algorithms to be biased towards the frequent records, and thus prevents them from learning infrequent records, which are usually more harmful to networks such as U2R and R2L attacks. It is possible to access the complete dataset here. The training dataset utilized in the NSL-KDD framework is referred to as KDDTrain+, while the corresponding test dataset is denoted as KDDTest+. The NSL-KDD dataset consists of records that contain a total of 42 attributes. There are a total of 41 attributes that represent the characteristic attributes of the data, while one attribute specifically represents the type of attack [41]. The training set KDDTrain+ of the NSL-KDD dataset has 125,973 network connection records. The test set KDDTest+ has 22,543 network connection records. The NSL-KDD dataset contains an overall total of 148,517 individual samples [40]. According to the data presented in Tables 3 and 4, there are a total of 23 separate attack types that have been classified into five distinct categories. The attack types are classified into four distinct categories, namely Denial of Service (DoS), Probe, User to Root (U2R), and Remote to Local (R2L). The NSL-KDD dataset shows a clear class imbalance, particularly between non-malicious traffic and different attack categories. Specifically, Denial of Service (DoS) attacks occur with high frequency, while categories such as U2R (User-to-Root) and R2L (Remote-to-Local) are severely underrepresented. This imbalance can lead to models' overfitting to dominant attack categories, potentially overlooking minority categories. To address this issue, oversampling was applied to reduce the dominance of dominant attack types, creating a more balanced training dataset. Model performance was then evaluated using metrics such as F1 Score, Precision, and Recall to ensure effectiveness across both majority and minority categories. Kindly refer to the citation of datasets https://www.unb.ca/cic/datasets/nsl.html.

**Table 3.** Characteristics of NSL-KDD datasets

| Classes | 37 types of attacks |
|---|---|
| DOS | Back, Mailbomb, Neptune, Pod,Smurf, Udpstor m, Teardrop, Processtable, Apache2, Worm |
| Probe | Satan, IPsweep, Portsweep, Mscan, Nmap,Sa int |
| R2L | Guess_password, Ftp_write,Imap, Warezmaster, Xlock,Phf, Xsnoop, Snmpgue, Snmpgetattack, Multi hop, Named Httptunnel, Sendmail, |
| U2R | Buffer_overflow, Rootkit,Perl ,Loadmodule, ,Sqlattack,Xterm,Ps |
| Normal | Normal |

**Table 4.** Classification types of features in NSL-KDD

| Features | Type | Features | Type | Features | Type |
|---|---|---|---|---|---|
| duration | int64 | num_file_creations | int64 | diff_srv_rate | float64 |
| protocol_type | int64 | num_shells | int64 | srv_diff_host_rate | float64 |
| service | **object** | num_access_files | int64 | dst_host_count | int64 |
| flag | **object** | num_outbound_cmds | int64 | dst_host_srv_count | int64 |
| src_bytes | **object** | is_host_login | int64 | dst_host_same_srv_rate | float64 |
| dst_bytes | int64 | is_guest_login | int64 | dst_host_diff_srv_rate | float64 |
| land | int64 | srv_count | int64 | dst_host_same_src_port_rate | float64 |
| wrong_fragment | int64 | serror_rate | float64 | dst_host_srv_diff_host_rate | float64 |
| urgent | int64 | srv_serror_rate | float64 | dst_host_serror_rate | float64 |
| hot | int64 | rerror_rate | float64 | dst_host_srv_serror_rate | float64 |
| num_failed_logins | int64 | srv_rerror_rate | float64 | dst_host_rerror_rate | float64 |
| logged_in | int64 | same_srv_rate | float64 | dst_host_srv_rerror_rate | float64 |
| num_compromised | int64 | su_attempted | float64 | class | **object** |
| 'root_shell' | int64 | num_root | int64 | - | - |

## 5 METHODOLOGY

Machine Learning (ML) methods combined with the Energy Valley Optimizer (EVO) for feature selection aimed to create a balanced dataset, leading to enhanced intrusion detection for wireless sensor networks (WSNs) and cloud computing. The validation of the proposed intrusion detection method included extensive testing and evaluation processes. The Energy Valley Optimizer (EVO), along with machine learning models, was run in Python using Google Colab Pro with an NVIDIA T4 GPU (High RAM) and a local machine equipped with an Intel Core i5 and 16GB RAM. The approach employed Scikit-learn for machine learning algorithms, along with NumPy and Pandas for data handling, and Matplotlib and Seaborn for data visualization. Custom Python scripts enabled EVO-based feature selection, ensuring quick execution and reproducibility.

### 5.1 Energy Valley Optimizer (EVO) for Feature Selection in Intrusion Detection Systems

This study employs an Intrusion Detection System (IDS) as a platform to evaluate the efficacy of our proposed feature selection method, the Energy Valley Optimizer (EVO). Intrusion Detection Systems (IDSs) are essential for detecting hostile actions in a network; yet, their precision and efficacy are frequently impeded by high-dimensional datasets including irrelevant and duplicate characteristics. Employ EVO as a feature selection technique to diminish dimensionality and improve IDS performance. Utilizing the NSL_KDD dataset, the influence of EVO for feature selection on the IDS's capacity to accurately differentiate between benign and malicious traffic while reducing computational overhead was assessed. Check Figures 5 and 6 of our proposed methodology on the IDS platform, which primarily encompasses the following procedures:

The initial phase preprocesses the NSL_KDD dataset by handling empty values, scaling data uniformly, and converting nominal features to numeric [23]. EVO-based feature selection is then applied, reducing irrelevant features to improve model efficiency and accuracy [23, 42]. The NSL-KDD dataset has 37 attack types classified into DoS, Probe, U2R, and R2L, exhibiting substantial imbalances; for instance, U2R attacks are markedly underrepresented relative to DoS. This disparity was rectified by downsampling to ensure a balanced distribution of training data [23, 43]. This disparity may result in biased models that overfit the majority class and underperform on the minority class. To resolve this issue, we implemented a downsampling strategy, decreasing the number of majority class instances (Dos assaults) to align with the minority class (U2R traffic).

This method guarantees a more equitable dataset, mitigating bias and enhancing the model's generalization capability. Downsampling diminishes the total size of the training dataset, although it improves the model's ability to identify both benign and malicious traffic more efficiently. Furthermore, to guarantee a thorough assessment, we employed F1 Score, Precision, and Recall, instead of depending exclusively on accuracy, to evaluate model performance across both classes. The findings indicated that downsampling enhanced performance equilibrium, improving the identification of the minority class while preserving strong predictions for the majority class. The NSL_KDD dataset comprises 42 numerical features and different attack types, with a prevalence of benign instances and DoS attacks balanced to improve IDS training [23, 43]. Following the preprocessing steps, the datasets are divided into two portions for machine learning algorithms. The initial 80% of the data is allocated for training the model, while the remaining 20% is set aside for testing [23]. Kindly refer to Algorithm 1 and Figure 3

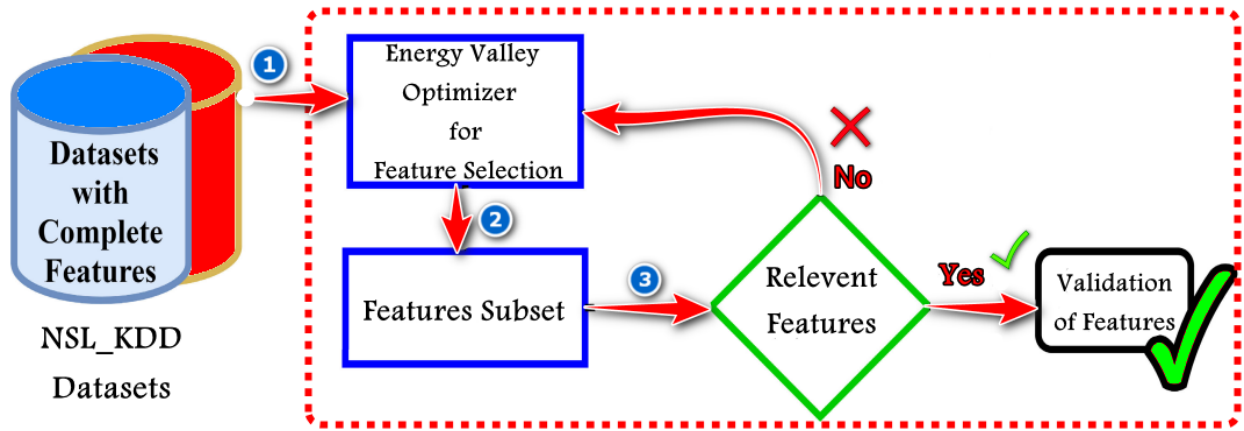

**Figure 3**. Framework of Proposed Novel Approach**:** Energy Valley Optimizer for Feature Selection

---
**Algorithm 1: Pseudocode of Energy Valley Optimizer (EVO) for Feature Selection**

---

**Input:** Train data: train = (Xi, Yi) // Training data
        MaxIter = Maximum number of iterations
        PopSize = Population size
**Output:** Best_features     // Optimal selected features
*# Initialize population and evaluate initial fitness*
1: population = initialize_population (PopSize)
2: fitness = evaluate_fitness (Population, Train_data)
*# Main optimization loop*
3: **for** iter = 1 to MaxIter do
4:    **for** each candidate in a population, do
5:        energy valley = calculate_energy (candidate)
6:        neighbors = find neighbors (candidate, energy_valley)
7        update_position (candidate, neighbors)
8:        candidate_fitness = evaluate_fitness (candidate, train_data)
9:    **end for**
10:   Best_candidate = select_Best (population, fitness)
11:   **if** stopping_criteria_met then
12**:**     **break**
13:   **end if**
14: **end for**
*# Return optimal feature subset*
15: **return** *Best_features = Best_candidate. features*

---

Where:
- (Xi, Yi): Training dataset, where Xi represents feature vectors and Yi denotes class labels.
- MaxIter: Maximum number of iterations.
- PopSize: Population size.
- P: Population of candidate feature subsets.
- Fitness: Fitness function to evaluate the quality of a feature subset.
- Neighboring: Function to find neighboring solutions based on the energy valley.
- Best_features: Optimal selected feature subset (best features for our target).
- EB: Energy Barrier
- BS: Best Particle
- nParticle: New Particle
- SL: Particle Slop
- X_CP: Center Point
- X_NG: Gravitational Point
- MaxFes: Maximum Feature

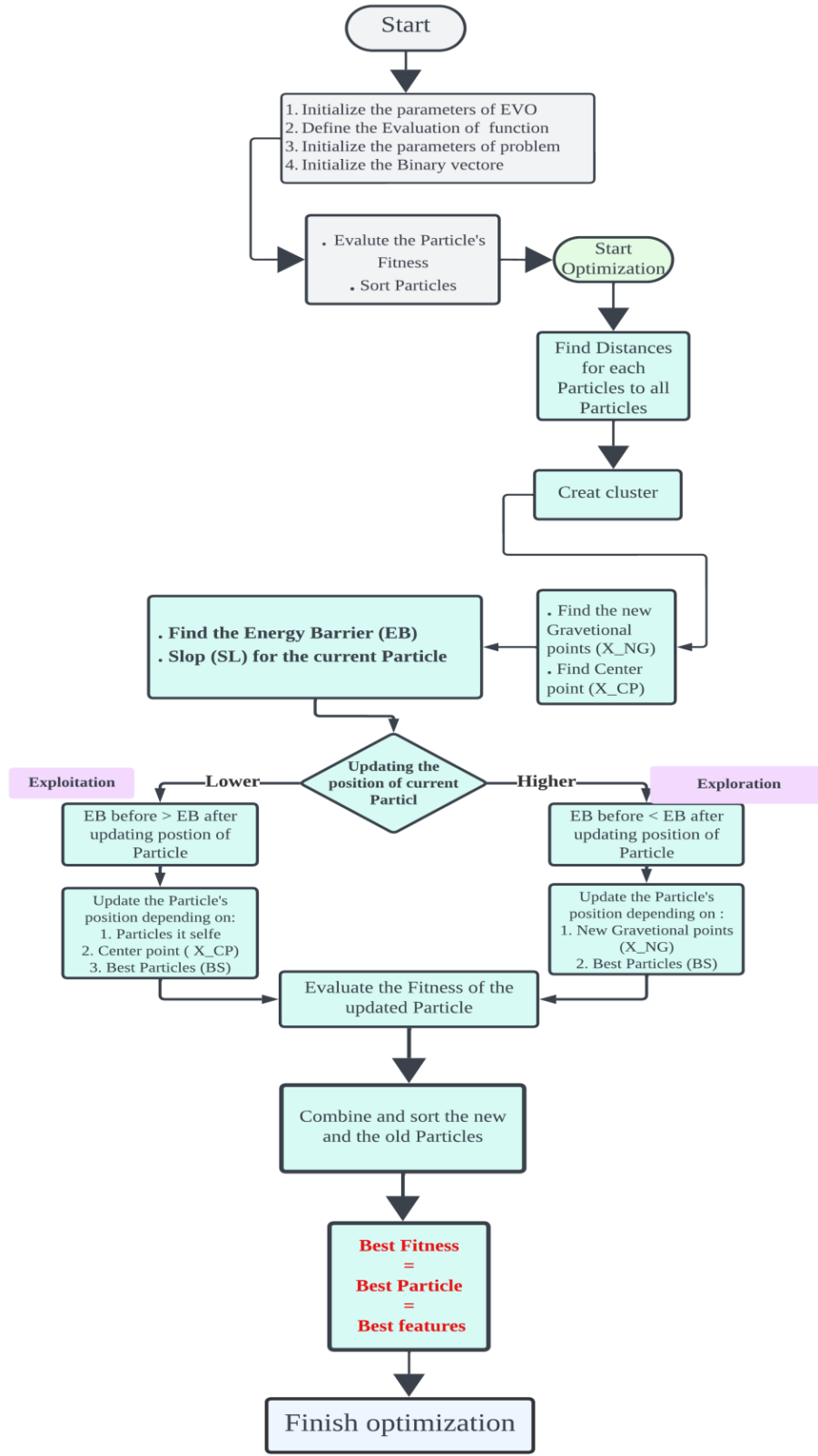

**Figure 5**. Flowchart of Proposed Novel Approach Energy Valley Optimizer for Feature Selection

## 5.2 Downsampling

This research tackles the problem of class imbalance present in the NSL-KDD dataset using a systematic downsampling method. The dataset includes five primary traffic categories: DoS (Denial of Service), Probe, R2L (Remote to Local), U2R (User to Root), and Normal. Notably, the DoS category contains a significantly higher number of instances (45,927) compared to the U2R and R2L categories, which are drastically underrepresented with only 52 and 995 instances, respectively. Such discrepancies can result in skewed intrusion detection models that are ineffective at identifying rare but critical attack types.

To address this issue, a class-based downsampling technique was applied to the training subset (KDDTrain+). The dataset was first divided by class labels. The dominant DoS class was downsampled [44] to equal the size of each minority class individually, thereby creating a balanced training dataset customized for each underrepresented category. Specifically, instances from the DoS class were resampled to match the counts of the U2R, R2L, Probe, and Normal traffic classes. The samples from minority classes were kept intact to maintain their statistical properties.

This balancing method guarantees equal representation of all classes during training, enhancing overall generalization and detection performance, especially for uncommon attack types. Model effectiveness was then assessed using metrics such as Precision, Recall, and F1-Score to gauge the success of the balancing technique across both majority and minority classes. Refer to Algorithm 2.

**Algorithm 2.** Procedure Balance for NSL-KDD Training Dataset

**Input:** An Imbalanced training dataset from NSL-KDD (KDDTrain+)
**Output:** A Balanced training dataset

1. **Initialize** an empty dataset: balanced_training_dataset
2. **Define** the majority class as "DoS" with 45,927 samples
3. **Identify** minority classes:
4. • Normal (13,449 instances)
5. • Probe (11,656 instances)
6. • R2L (995 instances)
7. • U2R (52 instances)
8. **Apply** downsampling to the majority class "DoS" to match
9. largest minority class "Normal" (13,449 instances)
10. **for** each class `C` in {Normal, Probe, R2L, U2R} do
11.     **if** instance count of class `C` < 1,000 then
12. Downsample the majority class "DoS" to match instance count of `C`
13.         add downsampled "DoS" data and class `C` to balanced_training_dataset
14.     **else**
15.         add all samples of class `C` to balanced_training_dataset
16.     **end if**
17. **end for**
18. **return** balanced_training_dataset
19. **end** procedure

## 5.3 Machine Learning Models

Machine learning classification is performed on the NSL_KDD datasets utilizing Logistic Regression (LogReg), Decision Tree (D_Tree), and K-Nearest Neighbors (KNN) methods to classify data as benign or malicious. A decision tree is a type of supervised machine learning algorithm employed for classification and regression problems, notably in classification tasks [45].

Noteworthy is its ability to handle both numeric (continuous) and nominal (categorical/attributes) data, considered a significant advantage [46]. A Decision Tree comprises two types of nodes: Decision Nodes, which facilitate choices with numerous branches, and Leaf Nodes, which represent the outcomes of decisions without additional branches [47]. KNN stands out as one of the simplest machine learning algorithms [48], primarily utilized for classification tasks, though capable of regression as well. KNN classifies new data by assuming similarity with existing data points, determining categories based on the highest degree of resemblance [49]. Logistic regression predicts the probability of events, such as stealth attacks, by classifying logs into normal or attack categories. It uses independent variables to predict a categorical outcome and can also handle multiclass scenarios to estimate the likelihood of different outcomes [50]. Figure 4 demonstrates the proposed ML models.

Hybridizing these machine learning models with the evolutionary algorithm for feature selection. This methodology finds essential elements and improves the precision of cybersecurity, attaining maximal accuracy in the least amount of time [16, 39]. Upon completion of the loop, the global best position obtained from the optimization loop is utilized to train the final ML model.

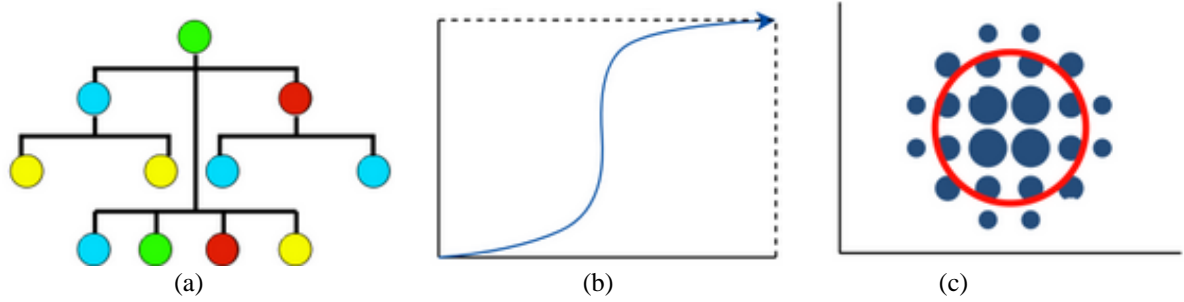

**Figure 4.** Proposed ML models: (a) Decision Tree, (b) Logistic Regression, (c) K-Nearest Neighbors

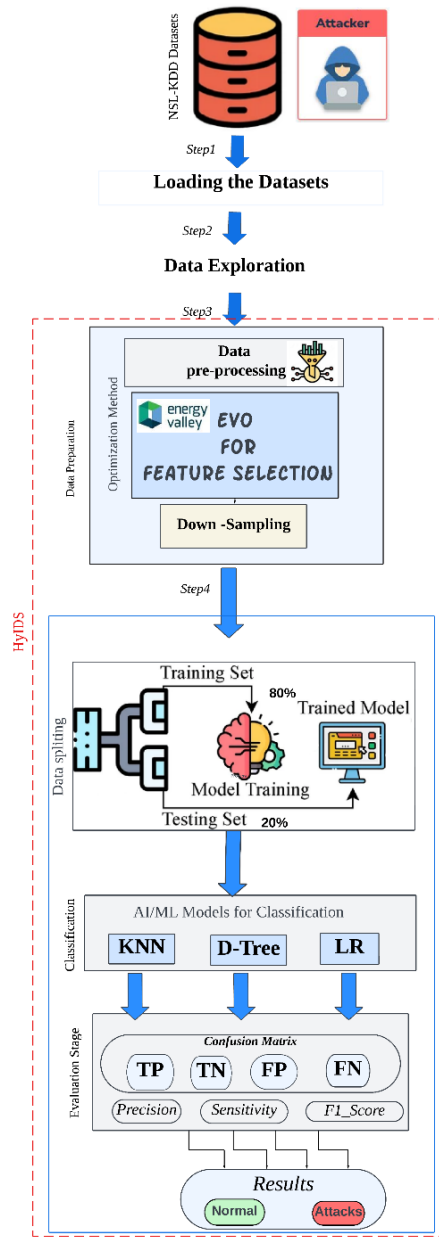

**Figure 6**. Flowchart of our proposed methodology (EVO for FS) on the IDS platform

# 6 RESULTS AND DISCUSSION

This section examines the evaluation of our novel proposal, EVO, for feature selection in the analysis of machine learning models for data classification utilizing the NSL-KDD dataset. The efficacy of machine learning models trained on data before and after optimization via our suggested technique is evaluated.

Table 5, Figures 7 and, present and summarize 8 accuracy results for each model and the classification results based on all 42 features before optimization, highlighting training and testing durations using the Decision Tree, KNN, and Logistic Regression (LogReg) models. Figure 9 illustrates a comparison of the durations for training and testing models before the implementation of any feature selection optimizer.

Table 6 and Figures 10, 11 and 12 analysis and outline the categorization outcomes for models employing features refined by the Energy Valley Optimizer (EVO) utilizing 38 features of the NSL_KDD dataset before feature selection optimization indicates that the Decision Tree (D_Tree) achieved an accuracy of 98.95%, demonstrating the quickest training time of 0.03367 seconds and testing time of 0.00581 seconds. K-Nearest Neighbors (KNN) attained an accuracy of 98.47%, although it exhibited the longest testing duration (0.0168 sec), hence diminishing its efficacy for real-time detection. Logistic Regression attained an accuracy of 88.84%, although it exhibited the longest testing duration (0.002103 sec), hence diminishing its efficacy for real-time detection.

**Table 5**. Classification results based on all features/before optimization (42 features)

| Model | Number of features | Accuracy | Precision | Recall | F1-score | Training Time(sec.) | Testing Time(sec.) |
|---|---|---|---|---|---|---|---|
| LogReg | | 87.94% | 88.67% | 88.48% | 88.41% | 16.08020 | 0.004958 |
| D_Tree | 42 | 87.79% | 9934% | 9935% | 9934% | 0.10229 | 0.005765 |
| KNN | | 97.38% | 9738% | 9737% | 9737% | 0.01731 | 70.2130 |

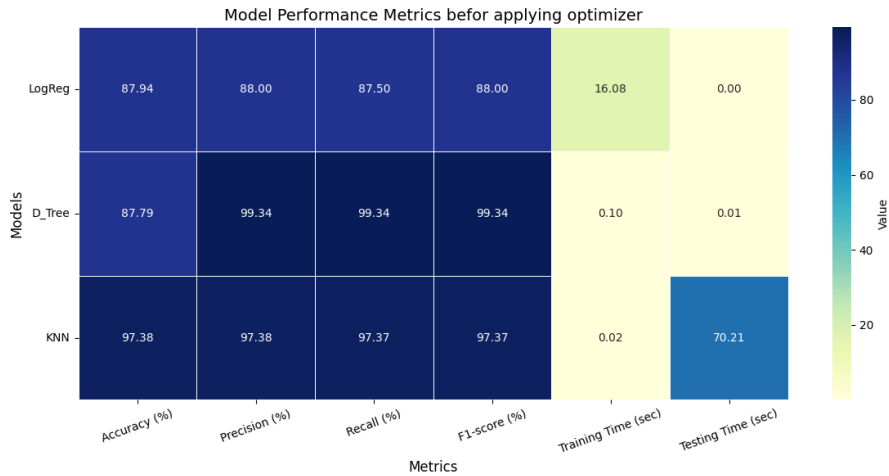

**Figure 7**. Performance of models before applying any FS optimizer

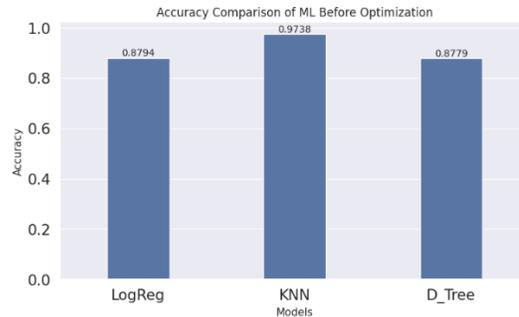

**Figure 8**. Comparison of models' accuracies before applying any FS optimizer

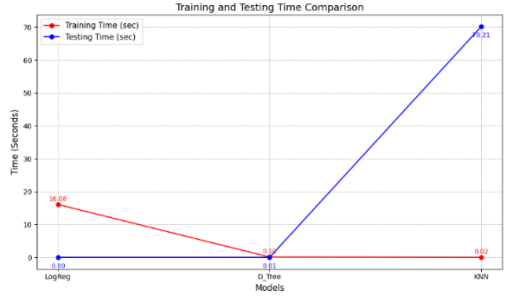

**Figure 9**. Comparison of train/test time lengths of models before applying any FS optimizer

**Table 6**. Performance improvement with EVO feature selection (18 features)

| Model | Number of features | Accuracy | Precision | Recall | F1-score | Training Time(sec.) | Testing time(sec.) |
|---|---|---|---|---|---|---|---|
| **LogRegEVO** |  | 88.84% | 89.00% | 89.00% | 88.50% | 78.79060 | 0.002103 |
| **D_TreeEVO** | **18** | 98.95% | 98.96% | 98.96% | 98.98% | 0.33679 | 0.005815 |
| **KNNEVO** |  | 98.47% | 98.47% | 98.47% | 98.47% | 0.016897 | 22.32870 |

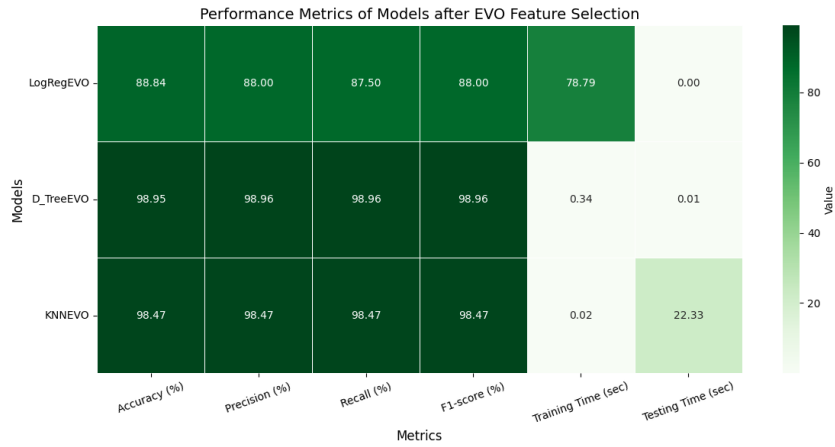

**Figure 10.** Comparison of time for models considering the application of EVO

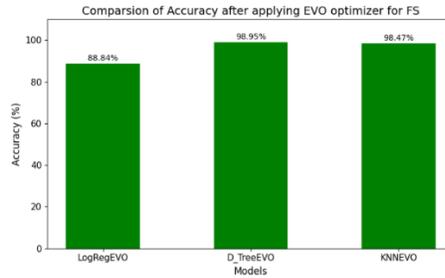

**Figure 11**. Comparison of accuracy after applying EVO

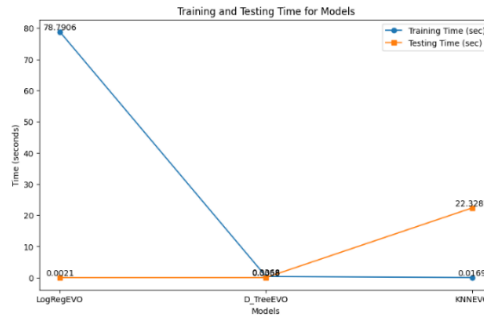

**Figure 12**. Comparison of train/test time lengths of models after applying the FS optimizer

**Table 7.** Comparison results with other methods based on the NSL-KDD datasets

| Authors | Datasets | Year | Proposed FS method | Accuracy |
|---|---|---|---|---|
| M Abdullah eta al. **[51]** | NSL_KDD | 2018 | IG-filters | 86.70% |
| S Sarvari et al. **[52]** | NSL_KDD | 2020 | MCF | 98.81% |
| H Alazzam, et al. **[53]** | NSL_KDD | 2020 | PIO | 88.30% |
| R. Zhao et al. **[23]** | NSL_KDD | 2022 | CFS-DE | 86.51% |
| H. Xu et al. **[54]** | NSL_KDD | 2023 | DL/BiLSTM | 79.54% |
| L Abualigah et al. **[55]** | NSL_KDD | 2025 | Improved chameleon swarm algorithm (ICSA) | 97.91% |
| *Our proposed method* | NSL_KDD | 2025 | KNN + EVOFS | **98.47%** |
| | | | D_Tree + EVOFS | **98.95%** |
| | | | LogReg + EVOFS | **88.84%** |

**Table 8.** Comparison results with other methods based on the feature selection method for multiple types of datasets

| Ref. | Year | FS Methods | Number of Features before optimization | Optimized Features |
|---|---|---|---|---|
| **Kumar et al. [56]** | 2021 | Tp2SF for Fs | 41 | 18 |
| **Nazir et al. [57]** | 2021 | TS-RF for FS | 43 | 16 |
| **SaiSindhu et al. [58]** | 2021 | Oppositional Crow Search Algorithm (OCSA) for FS | 41 | 10 |
| **Abdallah et al. [59]** | 2021 | Chi2-SMOST for FS | 84 | 20 |
| **Dey et al. [60]** | 2023 | Filter +NSGA-II for FS | 84 | 13 |
| **Dey et al. [60]** | 2023 | NSGA-II for FS | 84 | 18 |
| **Our proposed method** | 2025 | **EVO for FS** | **42** | **18** |

## 7 CONCLUSION

The research highlights the enhancement of the Energy Valley Optimizer for feature selection in large-scale real-world datasets. Our novel approach demonstrated strong performance and confirmed its effectiveness in enhancing the NSL_KDD dataset's attributes, with the feature set reduced while improving performance, namely, 18 out of 42 features increased. The enhanced dataset was assessed on an Intrusion Detection System (IDS) platform. EVO effectively reduces data dimensionality while improving classification performance, achieving an accuracy of Decision tree of 88.84%, precision of 89.00%, recall of 89.00%, and an F1 score of 88.50%, in additionally Logistic regression of 98.95%, 98.96%, 98.98% for Accuracy, precision, recall, and F1 score respectively. On the other hand, the KNN model shows an accuracy of 98.47% for each metric, utilizing only 18 features out of 42, while also optimizing the training duration in real-time. This development confirms the role of EVO in optimizing feature selection, significantly enhancing the performance of machine learning classifiers, such as Decision Trees, Logistic Regression, and KNN. The results demonstrate the effectiveness of EVO in enhancing real-world datasets, boosting intrusion detection capabilities, and protecting the cybersecurity systems of WSNs and cloud computing.

## 8 FUTURE WORK

Future research efforts will focus on improving the Energy Valley Optimizer (EVO) specifically for real-time Intrusion Detection System (IDS) environments. Real-time Intrusion Detection Systems require adaptive feature selection methods that are computationally efficient to handle high-velocity network traffic. The application of EVO in streaming environments enhances IDS responsiveness by reducing detection latency while maintaining high accuracy levels. The combination of the Energy Valley Optimizer (EVO) with another type of Optimizer and ensemble deep learning models that include Convolutional Neural Networks (CNNs), Recurrent Neural Networks (RNNs), and Long Short-Term Memory (LSTM) networks improves intrusion detection systems (IDS) by effectively recognizing both spatial and temporal attack patterns.